\documentclass[prc,aps]{revtex4}
\usepackage{graphicx}
\begin{document}

\title{Collapse of the random phase approximation: 
examples and counter-examples from the shell model}

\author{Calvin W. Johnson}
\affiliation{Department of Physics, San Diego State University,
5500 Campanile Drive, San Diego, CA 92182-1233}

\author{Ionel Stetcu}
\affiliation{Department of Physics, University of Washington, Box 351560, Seattle, WA 98195-1560}

\begin{abstract}
The Hartree-Fock approximation to the many-fermion problem can break
exact symmetries, and in some cases by changing a parameter in the
interaction one can drive the Hartree-Fock minimum from a
symmetry-breaking state to a symmetry-conserving state (also referred to 
as a ``phase transition'' in the literature). The order of
the transition is important when one applies the random phase
approximation (RPA) to the of the Hartree-Fock wavefunction: if
first order, RPA is stable through the transition, but if
second-order, then the RPA amplitudes become large and lead to
unphysical results. The latter is known as ``collapse'' of the RPA.
While the difference between first- and second-order transitions in
the RPA was first pointed out by Thouless, we present for the first
time non-trivial examples of both first- and second-order
transitions in a uniform model, the interacting shell-model, where
we can compare to exact numerical results.
\end{abstract}
\maketitle

\section{Introduction}

An important class of approximations in many-body
theory are mean-field approximations, which reduce the many-body
problem to an effective one-body problem, and their generalizations \cite{ring}. In this
paper we consider the Hartree-Fock approximation, a variational
approach that approximates the ground state wavefunction by a single
Slater determinant (antisymmeterized product of single-particle
wavefunctions), and the random phase approximation (RPA), which
builds small amplitude correlations on top of the Hartree-Fock state,
and which itself can be derived as the small-amplitude limit of the
\textit{time-dependent} Hartree-Fock approximation.

Because the Hartree-Fock (HF) state ignores correlations, it
can break exact symmetries, such as translational and
rotational invariance. A related approximation, the
Hartree-Fock-Bogoliubov (HFB) approximation, built upon
quasi-particles, also breaks conservation of the number of
particles. Despite this breaking of symmetries, HF, HFB, and the RPA and
quasi-particle RPA are widely and successfully used to describe
low-energy nuclear structure.

If one dials the parameters of the Hamiltonian, it is possible 
for the ground state HF or HFB wavefunction to
change from a symmetry-preserving state to one that breaks an exact
symmetry. For example, as one changes the single-particle energies as 
one attempts to fit data, the HF state may go between spherical and 
deformed states, and the HFB state may between normal
(number-conserving) and superfluid (number-nonconserving) states.

A long time ago Thouless pointed out \cite{Th61} that, as the mean-field state
is driven between symmetry-conserving (SC) and 
symmetry-nonconserving (SNC) states, there are two possible kinds of
state transitions (often referred to as ``phase transitions'' in the literature, 
even for finite systems). For second-order transitions the fluctuations in
RPA become unphysically large, leading to the so-called 
``collapse'' of the RPA. A frequently cited example is the collapse 
of RPA in the Lipkin-Meshkov-Glick model \cite{LMG65}. 
Significant effort in the literature has been devoted to
collapse in the RPA and QRPA and possible solutions. 
All but forgotten are first-order transitions for which RPA does 
\textit{not} collapse.

In this paper we use a non-trivial realization of the HF and RPA in
the interacting shell model,  where we can compare to exact
numerical calculations, and demonstrate both first- and second-order
transitions.  As predicted by Thouless, second-order transitions 
and collapse are driven by odd-parity operators or modes, 
while first-order transitions are driven by even-parity operators or modes. 
In this paper we do not propose any new solutions to the collapse of RPA at or near second-order transitions; instead, this work illustrates that the world of RPA and state transitions 
is more complicated than the standard narrative.

\section{Hartree-Fock and RPA calculations in the interacting shell
model}

We work in the framework of the interacting shell model.  The model
space is defined by a truncated set of single-particle orbits, for
example the  $0p_{1/2}$-$0p_{3/2}$ space, usually called the
\textit{p}-shell, or the $1s_{1/2}$-$0d_{3/2}$-$0d_{5/2}$ space, the
\textit{sd} shell. The interaction is given by a set of single-particle
energies $\epsilon_a$  and two-body matrix elements $V_{JT}(ab,cd)$.
The interactions used are not necessarily simple schematic forces but
generally start from carefully computed $G$-matrix interactions and
then adjusted empirically to reproduce a large number of ground
state binding energies and excitation energies. 

There are a number of computer programs that read in the model
space, generate many-body basis states (Slater determinants in
occupation space), compute the many-body matrix elements from the
single-particle energies and two-body matrix elements, and
diagonalize the Hamiltonian matrix to get out the eigenenergies and
wavefunctions, from which one can compute observables, transitions,
etc.. For this paper we used the REDSTICK shell model code\cite{REDSTICK}.

For Hartree-Fock plus RPA, we have use the SHERPA (SHEll-model RPA)
code \cite{SHERPA} which uses exactly the same input as REDSTICK.
SHERPA places no restriction on the
Hartree-Fock wavefunction except that it must be purely real;
otherwise it can have any arbitrary deformation contained in the
model space.  Thus one can have spherical, axially symmetric
deformation, triaxial deformation, and even parity-mixed ground
state.

In previous papers we have used SHERPA to directly test HF+RPA as an
approximation to full shell-model diagonalization, looking at
correlation energies\cite{stetcu2002}, ground state observables\cite{johnson2002}, electromagnetic
transitions\cite{stetcu2003}, and charge-changing Gamow-Teller transitions\cite{stetcu2004}.
In presenting those 
results we were frequenly asked about the issue of the collapse of RPA.
Such questions motivated this paper.

\section{First- and second-order transitions}

In this section we revisit Thouless' arguments \cite{Th60,Th61}
regarding the order of the transition. To do this, we need to remind
the reader of at least one way to develop RPA \cite{ring}, although
we do not give the derivation in full.

Let $| \Psi \rangle $ be a Slater determinant, that is, an antisymmeterized
product of  single-particle wavefunctions. In second quantization, where $\hat{a}^\dagger_a$
creates a particle in state $a$, we
write
\begin{equation}
| \Psi \rangle = \prod_i \hat{a}^\dagger_i | 0 \rangle.
\end{equation}
We follow the usual convention where $i,j$ denote occupied states
and $m,n$ denote unoccuped states. One then introduces the general
particle-hole operator, $\hat{Z}^\dagger = \sum_{mi} z_{mi}
\hat{a}^\dagger_m \hat{a}_i$, (where the $z_{mi}$ can be complex)
and the state
\begin{equation}
| z \rangle = \exp( \hat{Z} ) | \Psi \rangle
\end{equation}
which, by Thouless' theorem \cite{Th60} is a Slater determinant not orthogonal to
the starting state. One can compute
\begin{equation}
E(z) = \frac{\left \langle z \left | \hat{H} \right | z \right \rangle}{\left \langle z |  z \right \rangle}
\end{equation}
which, assuming $z$ small, can be expanded
\begin{equation}
E(z) = E_0 + \sum_{mi} h_{mi}^* z_{mi} + h_{mi} z_{mi}^* +
\sum_{mi,nj} A_{mi,nj}z_{mi}z_{nj}^* + \frac{1}{2}B_{mi,nj}z_{mi}^*
z_{nj}^* +\frac{1}{2}B_{mi,nj}^* z_{mi} z_{nj} + \ldots...
\end{equation}
At a local minimum, $h_{mi} = 0$; this is the Hartree-Fock condition,
and $E_0$ is the Hartree-Fock energy. The quadratic terms can be
treated as a harmonic oscillator: one treats the $z_{mi}$ as boson
operators and using a Bogoliubov transformation put into diagonal
form, using the famous RPA matrix equation:
\begin{equation}
\left ( \begin{array}{cc} \mathbf{A} & \mathbf{B} \\
-\mathbf{B}^* & -\mathbf{A}^*
\end{array} \right )
\left ( \begin{array}{c} \vec{X}_\lambda \\
\vec{Y}_\lambda \end{array} \right ) = \Omega_\lambda \left ( \begin{array}{c} \vec{X}_\lambda \\
\vec{Y}_\lambda \end{array} \right ) \label{rpaeqn}
\end{equation}
There can be, of course, higher-order corrections in the energy
landscape beyond quadratic, which we will consider shortly.

Suppose one is in a SC state, e.g., a state of good angular
momentum, usually spherical, or parity. Then, solving Eq.~(\ref{rpaeqn})
one finds the $X,Y$ modes also have good symmetry. For
example, with our code SHERPA, if the HF state is spherically
symmetric, one gets out RPA modes that have good angular momentum
$J$, and one sees the appropriate $(2J+1)$ degeneracy in the RPA
frequencies $\Omega_\lambda$. (If, on the other hand, rotational
symmetry is broken but one has axial symmetry, then one can have
two-fold degeneracies, signaling RPA modes  that are time-reversed
of each other, or a single mode, which must be time-reversal even.
In the event of triaxiality, one has no degeneracies in the RPA
spectrum.)

To consider ``state transitions,'' from SC to SNC, one needs to look
at higher-order terms. Let $\hat{Z}_\lambda$ represent generically
the RPA modes; by expanding about the HF minimum, one expands the
energy landscape by $\langle HF | \hat{Z}_\lambda^n | HF \rangle$.
Because we are at a minimum, $n=1$ vanishes. $n=2$ yield the
curvature and the RPA frequencies. What about $n=3,4,\ldots$?

Now we get to the heart of Thouless' argument. If the RPA mode
$\hat{Z}_\lambda$ has odd parity, while the HF state has good
parity, then $\langle HF | \hat{Z}_\lambda^n | HF \rangle$ must
vanish for all odd $n$. On the other hand, if $\hat{Z}_\lambda$ has
even parity, then $\langle HF | \hat{Z}_\lambda^3 | HF \rangle$ can
be nonzero; for example, it is possible to couple three $J=2$
operators to total angular momentum zero.

\begin{figure}
\centering
\includegraphics*[scale=0.5]{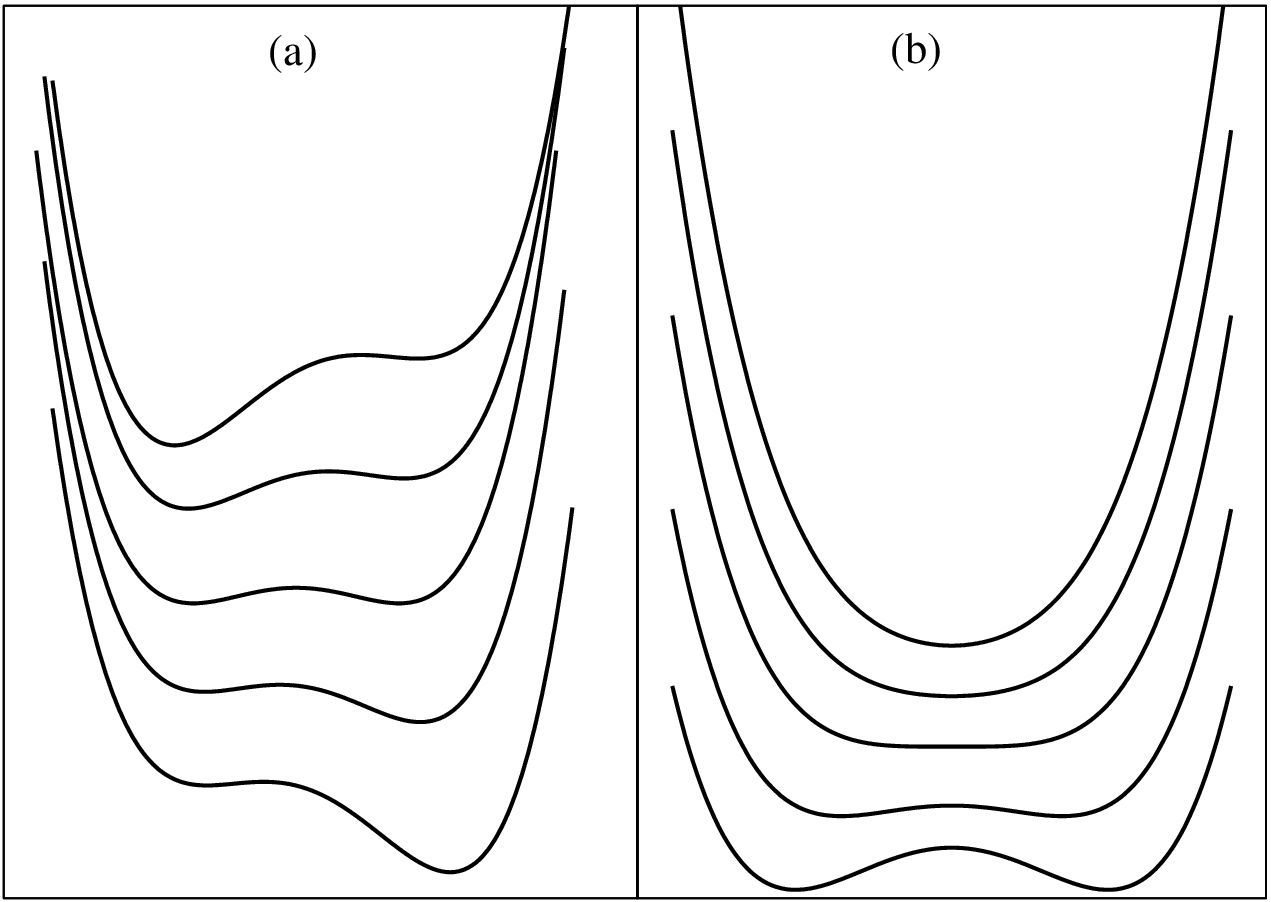}
\caption{Sketches of ``phase'' transitions in the energy landscape.
(a) A first-order transition. (b) A second-order transition.}
\label{pots}
\end{figure}

Figure \ref{pots} illustrates both cases. Fig.~\ref{pots}(a) is for
modes with even parity, so that cubic terms play a role. One can
clearly have coexisting local minima, and so one gets a first-order
transition. Fig.~\ref{pots}(b) is for odd-parity modes, so that the
energy landscape must be symmetric. One tends to get only a
second-order transition. 

In our examples below, we see Thouless' predictions played out. A
system with only even parity modes has first-order transitions,
while a system with odd parity modes has a second-order transition.

What happens in a second-order transition? In that case the RPA 
frequencies $\Omega_\lambda \rightarrow 0$ and the fluctuations about 
the HF state, measured by $|Y_\lambda|$, become very large, 
allowed because the RPA vectors have a 
nonstandard normalization: $|X_\lambda|^2 - |Y_\lambda|^2 = 1$.
The RPA energy is 
\begin{equation}
E_\mathrm{RPA} = E_\mathrm{HF} - \sum_{\lambda > 0} \hbar \Omega_\lambda 
|Y_\lambda|^2 - \frac{\langle P^2 \rangle }{2M_0}
\end{equation}
(see \cite{ring,stetcu2002} for details), and so as $|Y_\lambda|^2$ becomes 
large, the energy dives or collapses. The appearance of unphysical values 
is unsurprising because in the derivation one assumes small amplitudes for $X$ and $Y$. 

In the published literature, discussions regarding the behavior of
the RPA near a phase transition focus exclusively on
collapse of RPA \cite{ring,collapse}, that is, on second-order transitions.
Outside of Thouless \cite{Th61} there is no discussion of first-order 
transitions (indeed, the literature appears to uniformly refer to phase transitions 
without quantifying the order of the transition). 
Yet, as illustrated below, it is not hard to find a 
first-order transition.  It is possible that people using RPA have 
encountered first-order transitions without realizing it, as there is 
no catastrophic collapse to signal the transition. While clearly second-order 
transitions are more problematic, we find it instructive to explore 
both kinds.

\section{Results}

\subsection{Example of a first-order transition: deformation}

We begin with two case studies in the $sd$-shell, using Wildenthal's
universal $sd$ (USD) interaction\cite{Wildenthal}. The first of
these is  $^{28}$Si (6 valence protons and 6 valence neutrons). By
increasing the difference $\Delta$ between the $0d_{5/2}$
single-particle energy and the $1s_{1/2}$-$0d_{3/2}$ single-particle
energies, we can force the Hartree-Fock state to go from an oblate
deformed state to a spherical state with the $0d_{5/2}$ shell
filled. For convenience $\Delta=0$ corresponds to the original
Wildenthal values.

\begin{figure}
\centering
\includegraphics*[scale=0.5]{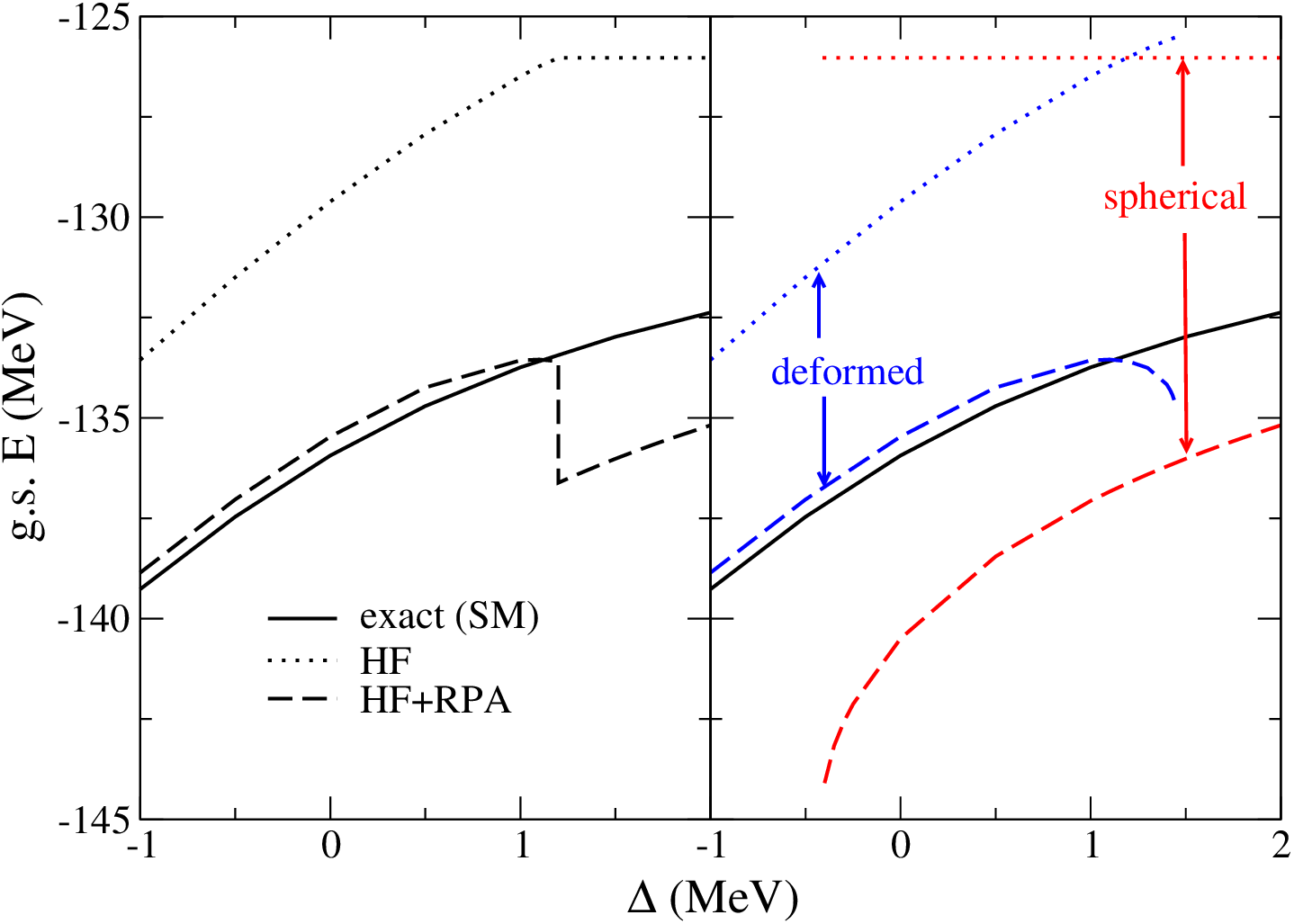}
\caption{(Color online) Ground state energy of $^{28}$Si, calculated in the full
interacting shell model (SM, solid lines), Hartree-Fock (HF, dotted lines), 
and Hartree-Fock plus random phase approximation (HF+RPA, dashed lines). Here $\Delta$ is added to the
$1s_{1/2}$ and $0d_{3/2}$ single-particle energies; $\Delta = 0$
corresponds to the original Wildenthal values. \label{si28gs}
The left-hand panel shows only the final result; the right-hand 
panel identifies spherical and (oblate) deformed mean-field phases. 
Note the significant region of coexistence.
}
\end{figure}

Fig.~(\ref{si28gs}) shows the ``exact'' calculation, which is an
interacting shell-model calculation performed in the full
$0\hbar\omega$ $sd$ valence space, compared to the lowest
Hartree-Fock energy, and the RPA correlation correction on top of
the HF energy. Because one switches between two degenerate 
HF states, the HF energy is continuous, while the HF+RPA energy 
is discontinuous, because the curvatures (RPA frequencies) are different. 
The right-hand panel illuminates in detail what is going on. 
Here we explicitly
show the (oblate) deformed and spherical HF energies and their respective RPA
corrections. One sees there is a significant region of
parameter space, encompassing the original Wildenthal value, where
locally stable deformed and spherical HF solution coexist.

\begin{figure}
\centering
\includegraphics*[scale=0.5]{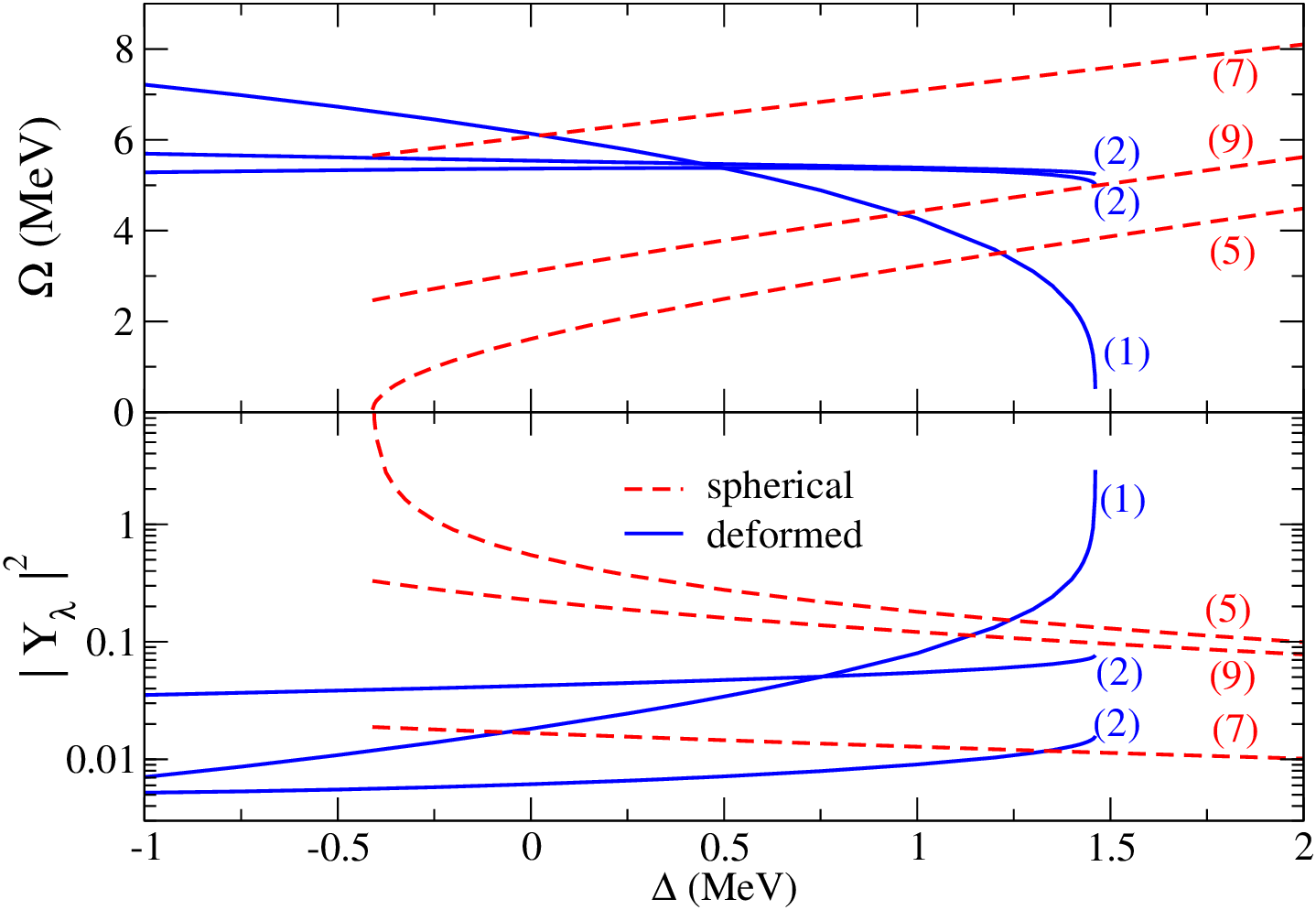}
\caption{(Color online) For $^{28}$Si. Upper panel: low-lying RPA
frequencies $\Omega$ for spherical (dashed) and deformed (solid) HF
states. Lower panel: $|Y_\lambda|^2$ corresponding to the 
RPA frequencies in the upper panel. In both cases the 
degeneracy is given in parentheses; for the spherical case 
the degeneracy  = $2J+1$ where $J$ is the angular momentum of 
the RPA mode.  \label{si28freq}}
\end{figure}

As we drive $\Delta$ further positive or negative, eventually the
deformed or spherical solutions, respectively, become unstable. In
Fig.~(\ref{si28gs}) this is seen as the HF+RPA energy dives sharply. 
For further diagnosis, in the upper half of
Fig.~(\ref{si28freq}) we plot the lowest nonzero RPA
eigenfrequencies $\Omega$ as a function of $\Delta$. (The deformed 
state also has two zero-frequency modes corresponding to broken
rotational symmetries.) We see that one eigenfrequency dives to
zero, signalling an instability. As further diagnosis, we 
give the degeneracy of the RPA eigenfrequencies; for the spherical HF 
state, the degeneracy of the collapsing eigenfrequency is 5, suggesting 
a quadrupole mode.  For the oblate deformed state, the RPA eigenmodes 
either come in time-reversed pairs (degeneracy = 2) or are already 
time-reversal-even (degeneracy = 1).

Note: We cannot follow the RPA frequency
all the way to zero, due to numerical instabilities,
although the trend is clear. Though we do not plot it, we also get a
corresponding eigenvalue of the the stability matrix diving to zero.

As the RPA frequency dives to zero, the corresponding magnitude of
the hole-particle amplitude, $|Y_{\lambda}|^2 = \sum_{mi}|Y_{mi,\lambda}|^2$, increases
dramatically.  This we plot in the lower half of
Fig.~(\ref{si28freq}). As discussed in a previous section, it is
this increase in $|Y|$ which causes the correlation energy to take
on unphysically large values. Although we do not show it, we have 
also computed the expectation value for various operators, such as 
the $Q \cdot Q$ operator; while the HF contribution is stable, the 
RPA correlation correction \cite{johnson2002} also shows 
unphysically large values.

\begin{figure}
\centering
\includegraphics*[scale=0.5]{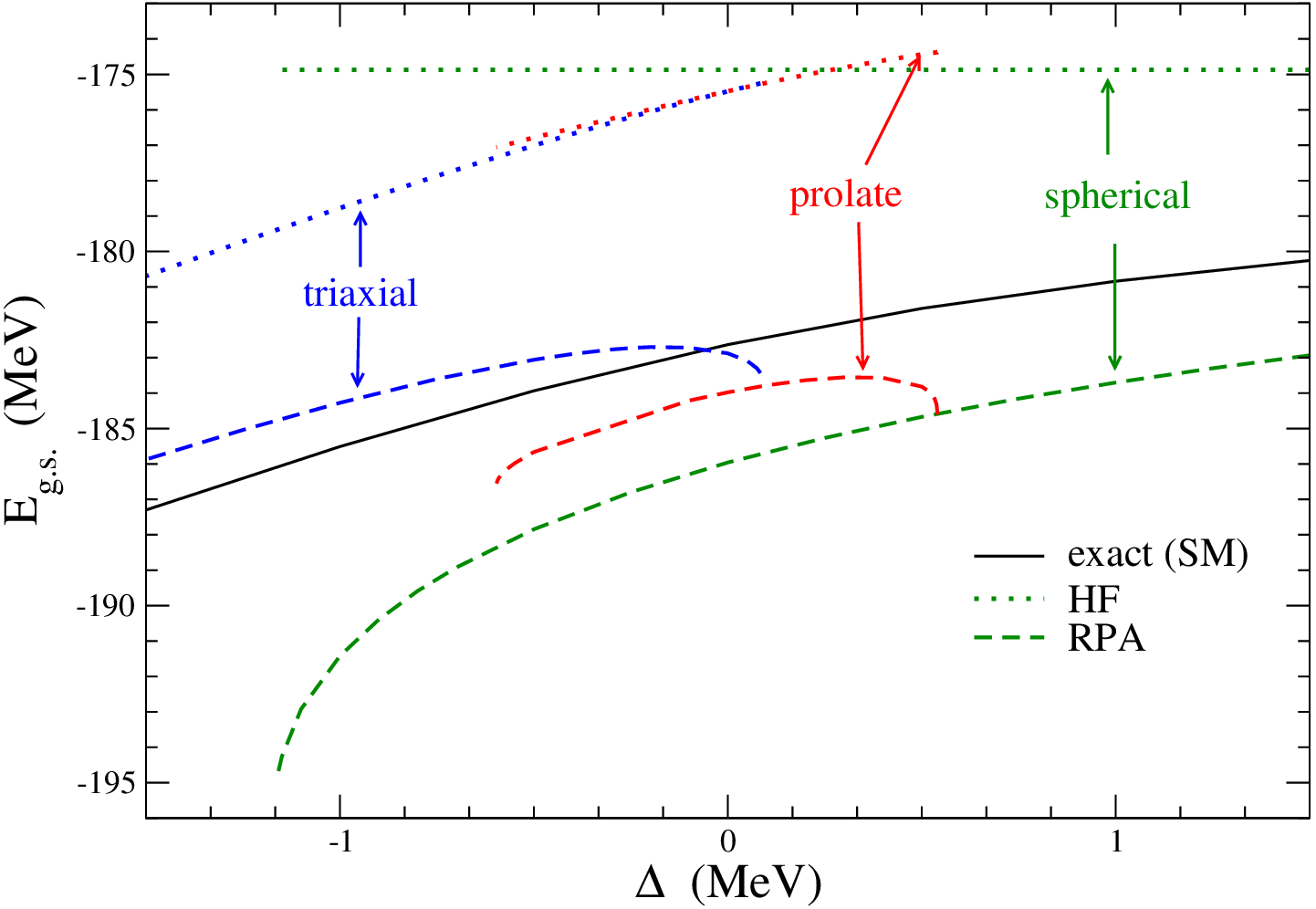}
\caption{(Color online) Similar to Fig.~(\ref{si28gs}) but for
$^{32}$S; here $\Delta$ is the change in the $0d_{3/2}$ energy. 
\label{s32gs}}
\end{figure}

\begin{figure}
\centering
\includegraphics*[scale=0.5]{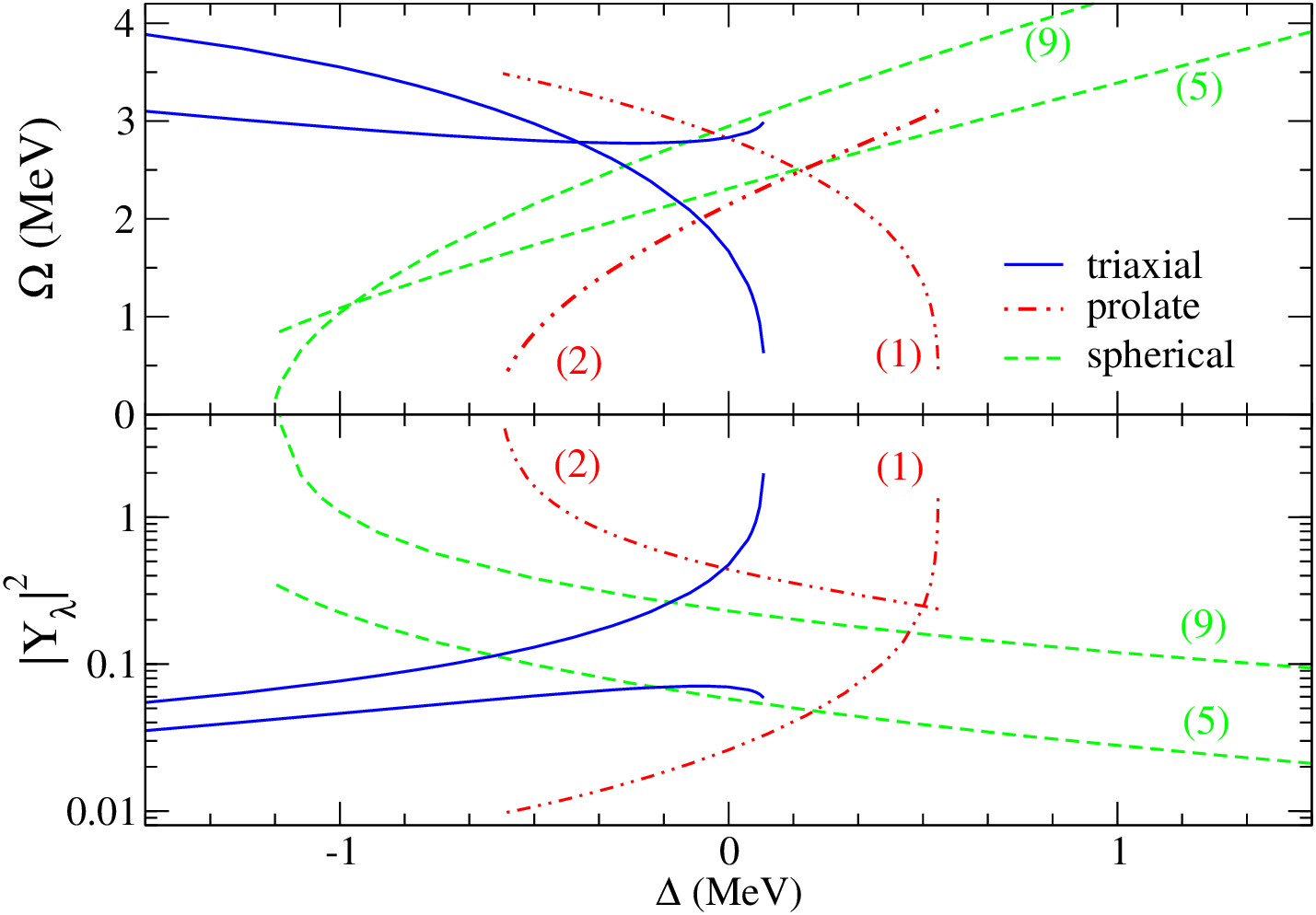}
\caption{(Color online) Similar to Fig.~(\ref{si28freq}) but for
$^{32}$S; here $\Delta$ is the change in the $0d_{3/2}$ energy.
\label{s32freq}}
\end{figure}

We also considered $^{32}$S, that is, 8 valence protons and 8
valence neutrons, and drove the $0d_{3/2}$ single-particle energy up
and down. This case was very interesting because we could get, by
adjusting $\Delta$, spherical, prolate, and triaxial solutions. 
Fig.~(\ref{s32gs}) shows the shell-model (SM), Hartree-Fock, and 
HF+RPA energies for the ground state, while Fig.~(\ref{s32freq}) 
shows the RPA frequencies and $|Y_\lambda|^2$. While a $J$=2 
(degeneracy = 5) eigenfrequency is falling, a $J$=4 (degenercy = 9) 
mode actually falls below it, suggesting that ultimately it is a 
hexadecupole mode becomes unstable first. The prolate state has both 
degenerate (time-reversed pairs) and non-degenerate states; the 
triaxial state, which has broken all possible symmetries, has no 
degenerate RPA modes.

We reiterate: the key idea here is that the coexistence of locally
stable phases leads to a first-order transition. The stable phases 
coexist easily because the parity-conserving mode--here quadrupole--means 
one can have cubic as well as quartic terms in the energy landscape. 
When we have an odd-parity mode, as illustrated in the next session, 
cubic terms are suppressed and one gets a second-order transition. 

\subsection{Example of a second-order transition: parity-mixing}

Second-order transitions lead to collapse of RPA. The classic example 
is the Lipkin model, which in its original form has a 
conserved parity (the Lipkin model is a two-level system, and the 
original Lipkin interaction could only promote or demote two particles 
at a time, thus providing an parity-like symmetry: either an even or 
an odd number of particles in the upper level); the transition of the 
HF state in the Lipkin model was from an exact parity state, with 
all particles in the lower level, to a mixed parity state. 

For our examples of second-order transitions, we consider a case involving shells of
opposite parity. We fix the $0s$ shell to be a closed core and have
as the valence space $0p_{3/2}, 0p_{1/2}$ and $0d_{5/2},
1s_{1/2}$.  We look at $^{12}$C, with 4 valence protons and 4 valence
neutrons, without any truncations on the many-body space. (The only
reason we leave out the $0d_{3/2}$ orbit is to make full shell-model
calculations tractable; SHERPA can easily handle the full $p$-$sd$
space, but the HF+RPA results look very similar to what we present
here. We also have looked at $^{16}$O and $^{20}$Ne in similar spaces and get
similar results.) We use the Cohen-Kurath (CK) matrix elements in
the $0p$ shell\cite{CK65}, the USD interaction \cite{Wildenthal} in
the $0d_{5/2}$-$1s_{1/2}$ space, and the Millener-Kurath (MK)
$p$-$sd$ cross-shell matrix elements\cite{MK75}.  Within the $p$ and
$sd$ spaces we use the original spacing of the single-particle
energies for the CK and USD interactions, respectively, but then
shift the $sd$ single-particle energies up or down relative to the
$p$-shell single particle energies by an amount $\Delta$; we define
$\Delta = 0$ where we get the first $3^-$ state at approximately
$6.1$ MeV above the ground state. The rest of the spectrum, in
particular the first excited $0^+$ state, is not very good, but the
idea is to have a non-trivial model, not exact reproduction of the
spectrum.

(As a side note, this model space is not translationally invariant, 
and so we do \textit{not} get zero-frequency modes from broken translational 
invariance.)

\begin{figure}
\centering
\includegraphics*[scale=0.5]{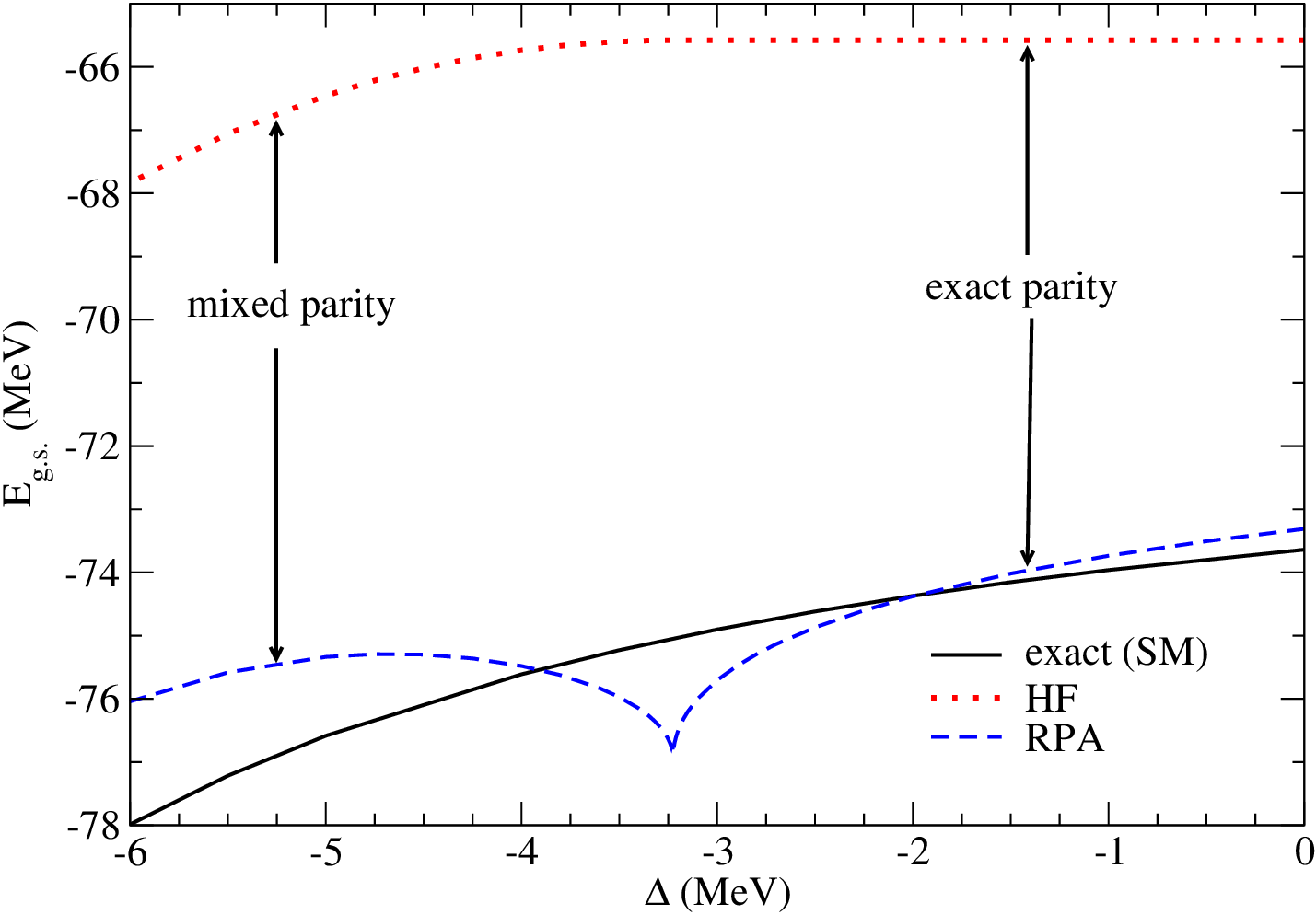}
\caption{(Color online) Ground state energy of $^{12}$C in the
$0p_{1/2}$-$0p_{1/2}$-$0d_{5/2}$-$1s_{1/2}$ space, calculated in the
full interacting shell model (SM) (solid line), Hartree-Fock (HF) (dotted line), and
Hartree-Fock plus random phase approximation (RPA) (dashed line). Here $\Delta$ is
added to the $1s_{1/2}$ and $0d_{5/2}$ single-particle energies;
$\Delta = 0$ puts the negative parity states in the SM at
approximately the correct location. We show where the HF states and
corresponding HF+RPA states have either exact parity (and also
spherical symmetry) or mixed parity (and break rotational invariance
as well) \label{c12gs}}
\end{figure}

\begin{figure}
\centering
\includegraphics*[scale=0.5]{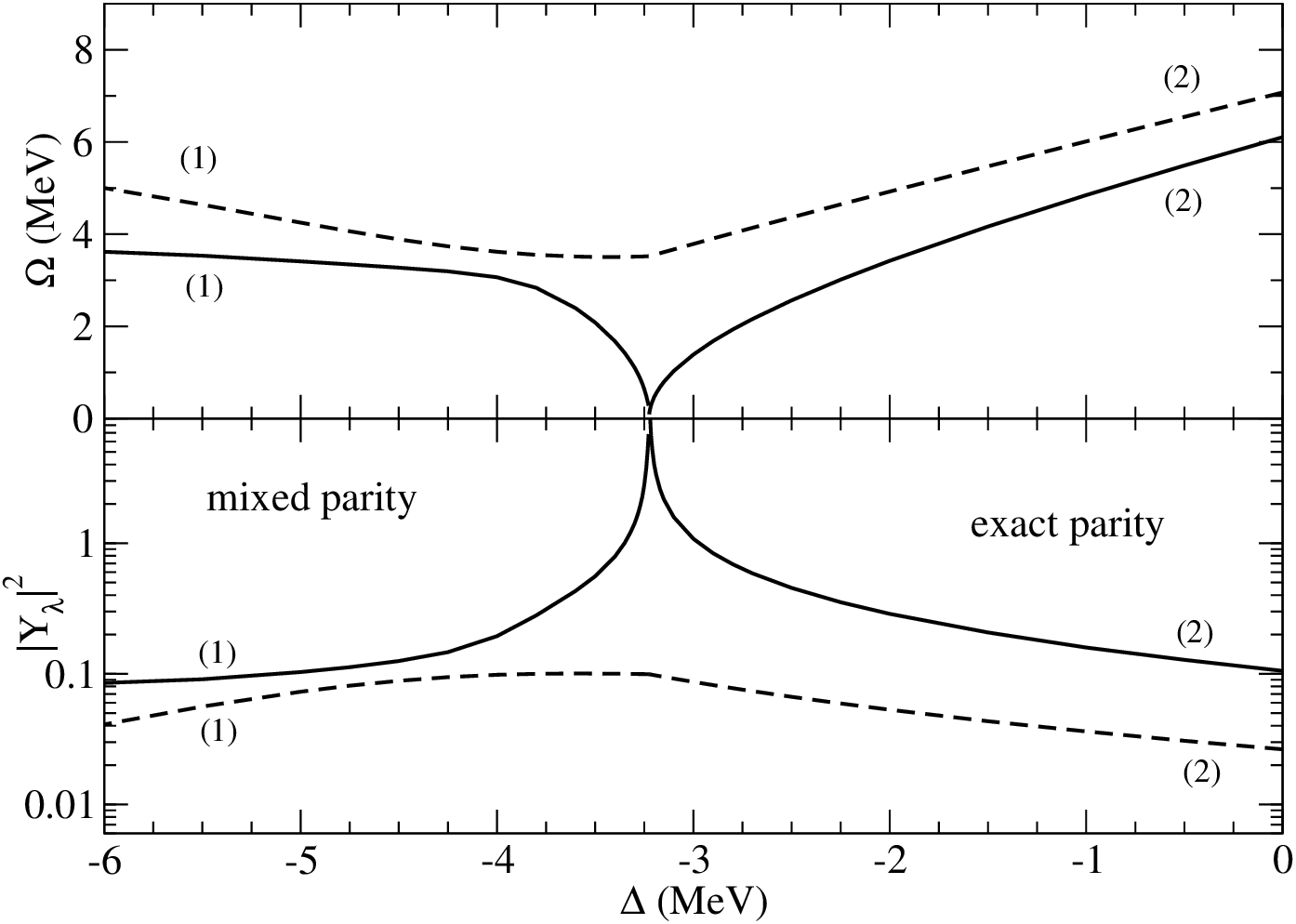}
\caption{For $^{12}$C. Upper panel: low-lying RPA
frequencies $\Omega$ for exact-parity and parity-mixing HF
states. Lower panel: $|Y_\lambda|^2$ corresponding to the 
RPA frequencies in the upper panel. In both cases the 
degeneracy is given in parentheses; the exact-parity  
HF states are oblate, allowing for time-reversed pairs, but for 
parity-mixing the HF state becomes triaxial, breaking the 
time-reversal degeneracy.
\label{c12freq}}
\end{figure}

Fig.~\ref{c12gs} compares the exact SM ground state energy with HF and HF+RPA. 
Here the HF+RPA energy dives, or ``collapses'' at the transition point.  
A more detailed look is in Fig.~\ref{c12freq} which plots the RPA frequencies. 

In Fig.~\ref{c12freq} we also show the RPA eigenfrequencies and the 
magnitude of $|Y_\lambda|$. The exact-parity HF states are oblate, 
so that the RPA frequencies come in degenerate pairs from time-reversal 
symmetry. The mixed-parity HF states are triaxial, breaking time-reversal 
degeneracy.

\section{Conclusions}

We have applied Hartree-Fock plus the random phase approximation in the 
framework of the interacting shell model, with complicated, realistic forces. 
By changing the single-particle energies we could drive the Hartree-Fock solution 
between symmetry conserving and symmetry non-conserving states. In accordance 
with Thouless' original, and oft-forgotten, analysis, first-order transitions 
are associated with even-parity modes, such as the quadrupole mode, 
and do not display, at the transition point, the infamous collapse of RPA. 
Instead one only obtains the collapse of RPA in second-order transitions, 
associated with odd-parity modes. The latter are of course more serious, but 
it is useful to keep in mind that not all state or phase transitions automatically 
lead to the collapse of RPA.

The U.S.~Department of Energy supported this investigation through
grants DE-FG02-96ER40985 and DE-FC02-07ER41457.

\end{document}